\documentclass[aps,11pt,onecolumn,superscriptaddress,showpacs,floatfix,nofootinbib, notitlepage]{revtex4-1}

\usepackage{amsmath,graphicx,color}
\usepackage{graphicx} 
\usepackage{booktabs}
\usepackage{multirow}
\usepackage{graphics}
\usepackage{hyperref}
\usepackage{bm}
\usepackage{calligra}
\usepackage[T1]{fontenc}
\usepackage{amssymb}
\usepackage{amsfonts}
\usepackage{mathrsfs}
\usepackage{dsfont}
\usepackage{footmisc}
\usepackage[normalem]{ulem}
\usepackage{xcolor}
\usepackage{lineno}
\usepackage{booktabs} 
\usepackage{siunitx}  

\graphicspath{{figuresc/}}

\newcommand{\media}[1]{\ensuremath{\left\langle #1 \right\rangle}}

\newcommand\TRENTo{T\kern-.4ex\raisebox{-.48ex}{R}\kern-.40ex ENTo}

\newcommand\GOUAIa{GOU\kern-.4ex\raisebox{-.45ex}{AI}\kern-.4ex}              
\newcommand\GOUAIc{G\kern-.35ex\raisebox{-.42ex}{\textsc{ouai}}\kern-.35ex}   
\newcommand\GOUAId{GOU\kern-.15ex\raisebox{.55ex}{\small AI}}                 

\begin{document}

\title{Final-state centrality in relativistic heavy-ion simulations from neural networks}

\author{Anibal Bezerra}
\affiliation{Departamento de Física, Instituto de Ciências Exatas, Universidade Federal de Alfenas, 37133-840, Alfenas, MG, Brazil
}

\author{Fernando G. Gardim}
\email{fernando.gardim@unifal-mg.edu.br}
\affiliation{Instituto de Ci\^encia e Tecnologia, Universidade Federal de Alfenas, 37715-400 Po\c cos de Caldas, MG, Brazil
}

\date{\today}

\begin{abstract}
Experiments in relativistic heavy-ion collisions determine centrality from final-state quantities, usually the charged-particle multiplicity. Performing the same in hydrodynamic simulations requires millions of events, which is impractical for state-of-the-art hybrid models, and centrality is therefore commonly approximated by the initial entropy or energy. We show that neural networks trained on a few hundred hybrid events predict the event-by-event multiplicity and mean transverse momentum $[p_T]$ with relative errors of 1--2\%, which makes it possible to generate millions of collisions at a small fraction of the cost of the full simulation. For Pb--Pb collisions at $2.76$~TeV and Au--Au collisions at $200$~GeV, the initial-state and final-state definitions select nearly the same events in 10\%-wide classes, but up to 20\% of the events change class when the classes are 1\% wide. We then compute the fluctuations of $[p_T]$ and its Pearson correlations with a proxy of the elliptic flow, estimated from the initial eccentricity, and with multiplicity. The first correlation is insensitive to how centrality is defined, while the second changes substantially. The approach can be applied at every design point of Bayesian analyses at modest additional cost.
\end{abstract}

\maketitle

\section{Introduction}

The discovery that the quark-gluon plasma (QGP) behaves as a nearly perfect liquid~\cite{ALICE:2022wpn, Busza:2018rrf, Heinz:2013th} motivated a computationally demanding program, turning final-state hadron spectra and correlations into quantitative constraints on the transport properties of the matter created at temperatures exceeding $10^{12}$~K~\cite{Gardim:2019xjs}. The theoretical framework is by now well established. Relativistic viscous hydrodynamics provides an effective description connecting the initial state of relativistic nuclear collisions to the freeze-out hypersurface at which hadrons are produced \cite{Gale:2013da, Romatschke:2017ejr}, and Bayesian inference has become the standard tool for extracting transport coefficients and initial-state properties from experimental data \cite{JETSCAPE:2020shq, Nijs:2020roc, Bernhard:2019bmu}. The main limitation is computational, a single hybrid event, from initial conditions through hydrodynamic evolution and hadronic transport, costs up to 4 CPU-hours in the boost-invariant setup used in this work, and training the emulators used in Bayesian analyses requires such simulations at hundreds of points in parameter space.

This limitation becomes more significant with the high-luminosity heavy-ion program of the LHC. Run~3 has already been completed, and Run~4 is scheduled to begin in 2030~\cite{CERN:LS3schedule}. With continuous readout, ALICE expects a minimum-bias Pb--Pb sample over Runs~3 and 4 roughly two orders of magnitude larger than that of Run~2~\cite{ALICE:2022wpn,Citron:2018lsq}. Centrality classes, defined as percentiles of the multiplicity measured by forward detectors, will then be determined with negligible statistical uncertainty even in classes much narrower than 1\%, and observables will be measured with high precision within them. A meaningful comparison with these data therefore requires simulations in which centrality is defined from an estimator as close as possible to the experimental one.

Defining centrality in this way is difficult in simulations. Bayesian analyses can afford only a limited number of events at each design point: the JETSCAPE Collaboration simulated 2500 minimum-bias events at each of 500 design points~\cite{JETSCAPE:2020mzn}, and the Duke group $\mathcal{O}(10^4)$~\cite{Bernhard:2016tnd}, in both cases defining centrality from the final multiplicity. These samples are orders of magnitude smaller than the experimental ones, and the resulting centrality classes carry large statistical fluctuations. A common alternative is to define centrality from the initial entropy or energy, which is strongly correlated with the final multiplicity and can be computed for $\mathcal{O}(10^5)$ or more initial conditions at negligible cost. This strategy is adopted by Trajectum~\cite{Nijs:2020roc} and by many studies at fixed model parameters~\cite{Shen:2015qta,Plumberg:2024leb,Gardim:2024nyz,Giacalone:2023cet}.

As we show below, the two definitions agree in wide classes but not in narrow ones. Narrow classes are needed for more than statistical precision. Within a wide class the impact parameter varies, and since most observables depend monotonically on centrality, this variation produces spurious correlations between them. Correlations should therefore be computed in 1\% classes and then recombined into wider ones~\cite{Gardim:2016nrr}, which requires a reliable centrality determination precisely where the initial-state and final-state estimators differ the most.

Neural networks (NN) provide a solution to this problem. Instead of speeding up a numerical solver for partial differential equations, they can learn the relationship between event-by-event initial conditions, such as entropy or energy density, and observables that can be experimentally measured, using a sampled training set. Once the network is trained, it becomes a surrogate model that can evaluate these observables significantly faster than a full simulation, enabling the generation of millions of events with minimal cost related to creating their initial conditions.

Several recent studies have explored this approach, utilizing NN to predict the anisotropic flow coefficients $v_n$, the charged-particle multiplicity $N_{\rm ch}$, and the mean transverse momentum $[p_T]$ from event-by-event initial conditions, trained on hybrid simulations employing relativistic viscous hydrodynamics \cite{Hirvonen:2023lqy, Hirvonen:2024ycx, Sun:2024lgo, OmanaKuttan:2025qjj, Stewart:2025vua, Auvinen:2026dyl}. These studies trained the networks on a dataset, consisting of a few thousand to tens of thousands of fully simulated events, which resulted in typical prediction errors of approximately 5\% \cite{Sun:2024lgo}.

This paper is organized as follows. In Sec. II, we outline how a NN trained on a limited set of hybrid events can predict the event-by-event global quantities, $N_{\rm ch}$ and $[p_T]$, for Pb--Pb collisions at $\sqrt{s_{NN}}=2.76$ TeV and Au--Au collisions at $\sqrt{s_{NN}}=200$ GeV. In Sec. III, we use the trained models to define centrality. In Sec. IV, we present the dependence of centrality on $[p_T]$ and its fluctuations at both energies. Section V discusses the correlation between $[p_T]$ and elliptic flow $v_2$, and with $N_{\rm ch}$. In Sec. VI we present our conclusions.

\section{Learning Multiplicity and Mean Transverse Momentum}
\label{NN}

Since the design points of Bayesian analyses typically contain hundreds to thousands of events, we train our networks on only a few hundred events of a full hybrid simulation. The goal is to obtain a fast and accurate event-by-event prediction of the $N_{\rm ch}$ and $[p_T]$, which in turn allows us to generate event samples large enough for the centrality classes to be determined with statistical precision comparable to that of experimental data. This enables a direct comparison between model and data for the global observables, which any hydrodynamic description must reproduce.

We use a state-of-the-art hybrid framework based on the Bayesian analysis of the JETSCAPE Collaboration~\cite{JETSCAPE:2020mzn}, with 17 model parameters fixed at their maximum a posteriori values for the Grad viscous correction, applied to Pb--Pb collisions at $\sqrt{s_{NN}} = 2.76$~TeV and Au--Au collisions at $\sqrt{s_{NN}} = 200$~GeV. The initial energy density is generated with \TRENTo~\cite{Moreland:2014oya} and followed by a free-streaming phase that brings the system toward local equilibrium~\cite{Broniowski:2008qk,Liu:2015nwa}. The viscous hydrodynamic evolution is performed with MUSIC~\cite{Schenke:2010nt,Paquet:2015lta} using a lattice QCD equation of state~\cite{HotQCD:2014kol}. At the end of the fluid phase, hadrons are sampled with iS3D~\cite{McNelis:2019auj}, and their subsequent decays and hadronic rescatterings are simulated with SMASH~\cite{SMASH:2016zqf}. Each hybrid event is particlized 100 times to sample the statistical fluctuations of iS3D and SMASH. The $N_{\rm ch}$ and $[p_T]$ used as training and validation targets are the average over these 100 oversamples. Particles are produced up to $0<p_T<5$ GeV and within $|\eta|<0.5$.

For each system, we simulated about 600 hybrid events minimum bias. Of these, 450 are used for training and the remaining 25\% form a test set. They are never used for training, early stopping, or model selection and serve only to quantify the accuracy of the final networks. Previous works trained on considerably larger samples, typically of order $10^4$ events~\cite{Hirvonen:2023lqy, Hirvonen:2024ycx, Sun:2024lgo,OmanaKuttan:2025qjj, Stewart:2025vua, Auvinen:2026dyl}, and in some cases used a separate network for each observable~\cite{Hirvonen:2023lqy,
Hirvonen:2024ycx}. Our approach differs in three respects. First, since an accurate event-by-event multiplicity is the key ingredient for defining centrality, we exploit the strong correlation between the total initial entropy $S$ and the final multiplicity. In addition to the initial-condition profile, $S$ is provided as an input to the network. Second, the smooth dependence of both observables on $S$ is not learned but fixed: each observable is written as $y = g(S)\,e^{d}$, where $g(S)$ is a trend fitted once to the training events and $d$ is the event-by-event deviation from it, which carries the information on the geometry and fluctuations of the initial state. The network is trained to predict only $d$. Since $g(S)$ alone captures more than 99\% of the variance of $\log N_{\rm ch}$, which spans three orders of magnitude, this prevents the trend from dominating the training and forces the network to learn the residual, which is precisely what distinguishes events of equal $S$ (see the Supplemental Material~\cite{SM}). Third, $N_{\rm ch}$ and $[p_T]$ are regressed simultaneously by a single network, with identical architecture and training procedure for both collision systems. Model selection uses five-fold cross-validation within the training events, and the final prediction is the geometric mean of the five resulting networks. The trained models are publicly available in the GOUAI (Global Observables Using Artificial Intelligence) repository~\cite{gouai}.

The initial condition can be given to the network either as entropy density, obtained from the \TRENTo\ energy density with the lattice equation of state~\cite{HotQCD:2014kol}, or directly as energy density. We trained identical networks, with the same training events and the same cross-validation folds, for both choices. Their accuracies are statistically indistinguishable, and their predictions for the same event differ by less than 1\% on average (see the Supplemental Material~\cite{SM}). Throughout this paper, we report results obtained with entropy density as input and observables computed from charged particles with $p_T \ge 0.15$~GeV, which matches the kinematic cut used by the ALICE Collaboration.

\begin{figure}[h]
    \centering
    \includegraphics[width=.8\linewidth]{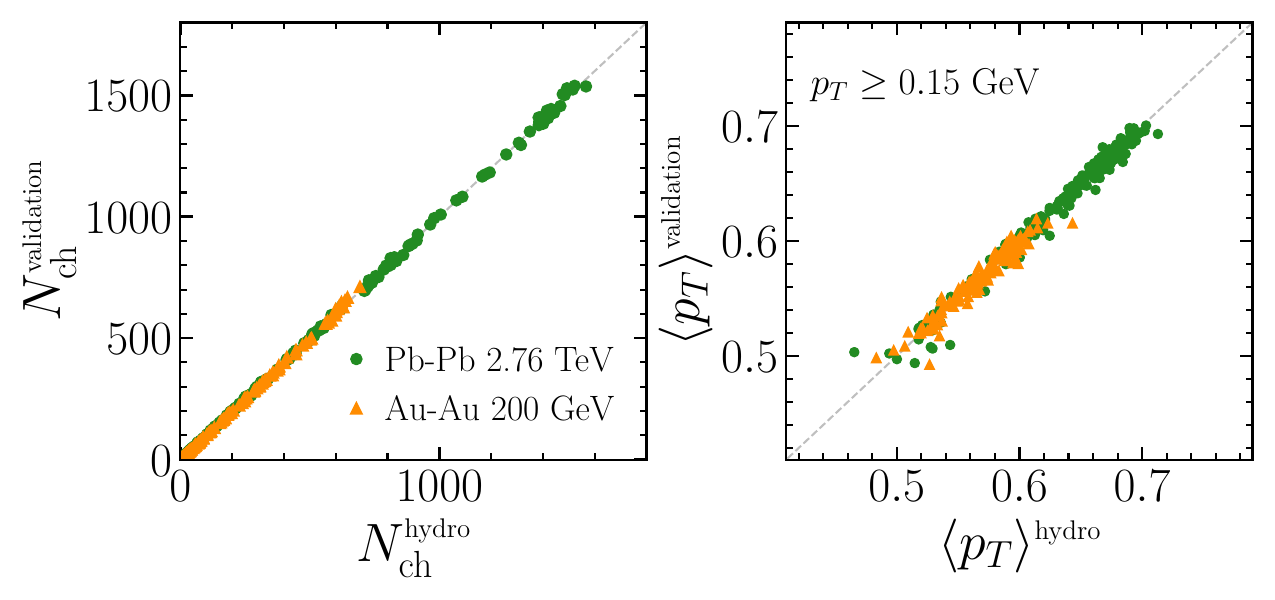}
    \caption{Validation of the neural network against full hybrid simulations
    for Au--Au collisions at $\sqrt{s_{NN}} = 200$~GeV (gold squares) and
    Pb--Pb collisions at $\sqrt{s_{NN}} = 2.76$~TeV (green circles), on the
    held-out events (135 and 169, respectively), which were not used for
    training or model selection. Entropy density is used as input and
    observables are computed with $p_T \ge 0.15$~GeV. Left: $N_{\rm ch}$
    predicted by the network versus the hybrid simulation. Right: the same
    for $[p_T]$.}
    \label{fig:validation}
\end{figure}

Figure~\ref{fig:validation} compares the network predictions with the full hybrid simulations for the held-out events. Up to 80\% centrality, the mean relative error is 0.5--1.8\% for $N_{\rm ch}$ and below 1.3\% for $[p_T]$ in both systems. Since the trend $g(S)$ alone accounts for most of the variance, the more stringent test is the recovery of the residual $d$, which the network reproduces with a correlation coefficient $r = 0.90$--$0.96$. In the most peripheral events, 80--100\%, where $N_{\rm ch} \lesssim 10$ and the training sample is sparse, the error of $N_{\rm ch}$ rises to 8--10\%, and results in these classes carry a larger model uncertainty. The architecture, training procedure, and further validation are given in the Supplemental Material~\cite{SM}.


\section{Impact on Centrality Definition}
\label{NN:centrality}

Experiments select events with final-state estimators, such as the multiplicity or the transverse energy measured in forward detectors. We now compare centrality classes defined from the total initial entropy with those defined from the charged-particle multiplicity predicted by the network, which we use as a final-state estimator, and examine how this choice affects the mean transverse momentum and its correlations.

The importance of a realistic centrality definition has been addressed previously in  \cite{Basak:2024fyq}, where a neural network trained on AMPT events was used to estimate the number of participants $N_{\rm part}$ in Au--Au collisions at $\sqrt{s_{NN}} = 200$~GeV from images of the initial
state. Here we extend this approach to a full hybrid description, with centrality defined from the final charged-particle multiplicity, and include the mean transverse momentum $[p_T]$.

For each system, we generated $1.5\times10^6$ \TRENTo\ initial conditions, with the same parameters as in the hybrid simulations~\cite{JETSCAPE:2020mzn}, converted them to entropy density with the EoS used in the hydrodynamics, and evaluated $N_{\rm ch}$ and $[p_T]$ with the trained networks. This took about 15 GPU-hours per system (under 2 hours of wall time on eight GPUs), including the generation of the initial conditions. For comparison, running the full hybrid chain for the same number of events would require of order $10^6$--$10^7$ CPU-hours.
 Results with the total initial energy are nearly identical.

\begin{figure}[h]
    \centering
    \includegraphics[width=1.0\linewidth]{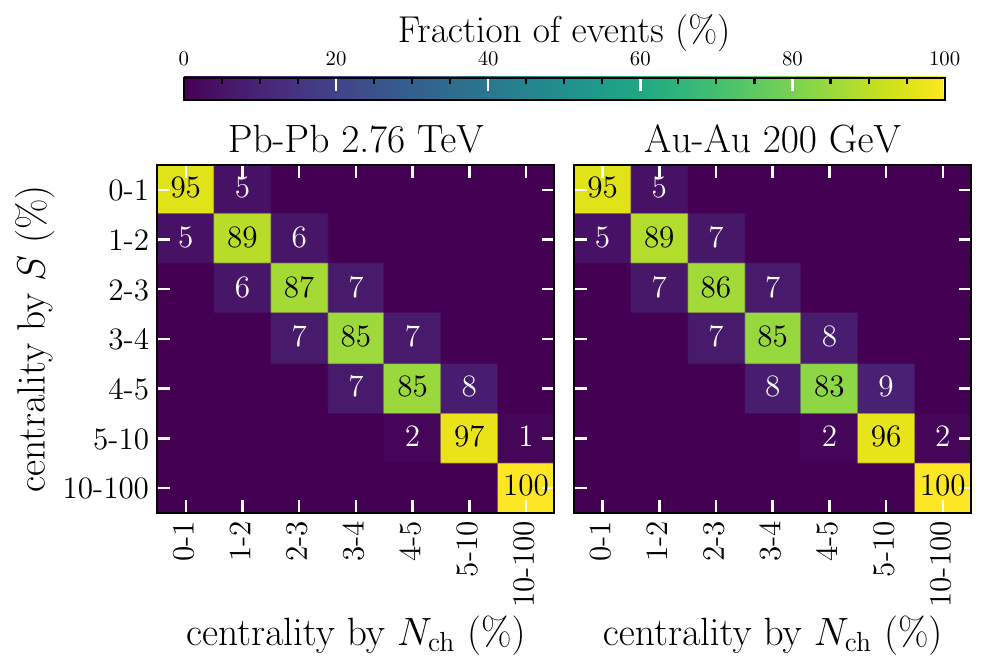}
    \caption{Migration matrices between centrality classes defined from the initial entropy $S$ and from the charged-particle multiplicity $N_{\rm ch}$ ($p_T \ge 0.15$~GeV) predicted by the neural network, for Pb--Pb collisions at $\sqrt{s_{NN}} = 2.76$~TeV (left) and Au--Au collisions at $\sqrt{s_{NN}} = 200$~GeV (right). Classes are 1\% wide up to 5\% centrality, followed by the 5--10\% and 10--100\% classes. Each entry gives the fraction of events in a given $S$ class that fall in each $N_{\rm ch}$ class.}
    \label{fig:centrality}
\end{figure}

The comparison between centrality classes defined from the initial entropy with those defined from the multiplicity $N_{\rm ch}$ predicted by the NN, for Pb--Pb collisions at $\sqrt{s_{NN}} = 2.76$~TeV (left) and Au--Au collisions at $\sqrt{s_{NN}} = 200$~GeV (right),  is presented in figure~\ref{fig:centrality} . The question is whether the events assigned to a class by the initial entropy are the same events selected by the multiplicity. In the 5--10\% class, 96--97\% of the events are assigned to the same class by both definitions. In 1\%-wide classes this fraction drops to 83--95\%, and the remaining events migrate only to the adjacent classes. The two systems behave almost identically.

As a cross-check that this is not an artifact of imperfect residual recovery, we also computed the same migration matrix directly from the $\sim600$ hybrid events, without the network: the diagonal agrees with the NN-based result to within 1--2 percentage points at 10\%-wide classes (98.7\% vs.\ 99.0\% for Pb--Pb, 96.9\% vs.\ 96.6\% for Au--Au). Since $r<1$ and $\beta>1$ for the residual (Supplemental Material~\cite{SM}), the predicted residual is under-dispersed by a factor $r^2\approx0.85$ in variance relative to the true one, which biases the predicted migration low; the agreement with the hybrid-only matrix indicates that this bias is small at 10\% width, though it is expected to matter more for the 1\%-wide classes, where it cannot be checked directly with only $\sim600$ hybrid events. 

The network is trained to predict the full event-by-event multiplicity, including the deviations from the smooth entropy trend that encode the geometry and fluctuations of the initial state (see the Supplemental Material~\cite{SM}), but it does not include the additional fluctuations from particlization and hadronic rescattering: each hybrid event used for training is itself the average of 100 particlization oversamples (Sec.~\ref{NN}), so these fluctuations are already suppressed in the training target and are not part of what the network learns. Finite-multiplicity fluctuations in a real experimental estimator would weaken the correlation between $N_{\rm ch}$ and the initial entropy still more, so the differences between the two centrality definitions shown here should be regarded as a lower bound. These results show that comparing a model with data requires the model's event selection to emulate the experimental estimator as closely as possible; with NN such as GOUAI, a multiplicity-based selection much closer to experiment than initial-state proxies is now computationally feasible.

We stress that centrality is a convention for event selection rather than an observable, and that different definitions -- impact parameter, number of participants, initial entropy or energy, and the various final-state estimators used by different experiments -- select different event ensembles, each appropriate for a different purpose. Our results do not imply that initial-state definitions are incorrect: they remain natural for studies of the hydrodynamic response at fixed initial conditions, and we find them equivalent to multiplicity-based centrality for classes 10\% wide or wider and for observables uncorrelated with the selection variable, such as $\rho(v_2^2,[p_T])$ (Sec.~\ref{rho}). What our results show is that, when model predictions are confronted with data, the event selection in the model should emulate the experimental estimator, and that this becomes essential for narrow classes and for observables that involve the multiplicity itself. The multiplicity-based estimator used here, the deterministic mean $N_{\rm ch}$ at midrapidity with $p_T\ge0.15$~GeV, is one more such estimator: it is closer to the experimental ones than $S$, since it shares their final-state, finite-acceptance character, but it is not identical to any of them, since it lacks both the finite-multiplicity resolution and the specific acceptance (e.g.\ ALICE V0M, STAR refmult) of a real detector.

\section{Transverse Momentum and Its Fluctuations}

We use the network trained with the initial entropy density as input -- shown in the Supplemental Material~\cite{SM} to be statistically equivalent to energy density -- and define centrality from the predicted charged-particle multiplicity $N_{\rm ch}$ with $p_T \ge 0.15$~GeV and $|\eta|<0.5$. This cut corresponds to the one used by ALICE in Pb--Pb collisions at
$\sqrt{s_{NN}} = 2.76$~TeV. We apply the same cut to Au--Au collisions at $\sqrt{s_{NN}} = 200$~GeV: since, beyond the centrality definition, we aim to isolate the effect of collision energy on the QGP, we keep the model parameters and the kinematic ranges identical for both systems, so that differences between them reflect physics rather than analysis choices.

As shown in Sec.~\ref{NN}, the network also predicts the event-by-event $[p_T]$, which allows us to study the momentum properties of the model with high statistics. In each centrality class containing $N$ events, we compute the average transverse momentum and its standard deviation,

\begin{align}
    \media{p_T} &= \frac{1}{N}\sum_{i=1}^{N}[p_T]_i, \\
    \sigma_{[p_T]} &= \sqrt{\frac{1}{N-1}\sum_{i=1}^{N}\left([p_T]_i-\media{p_T}\right)^2}.
\end{align}

\begin{figure}[h]
    \centering
    \includegraphics[width=.8\linewidth]{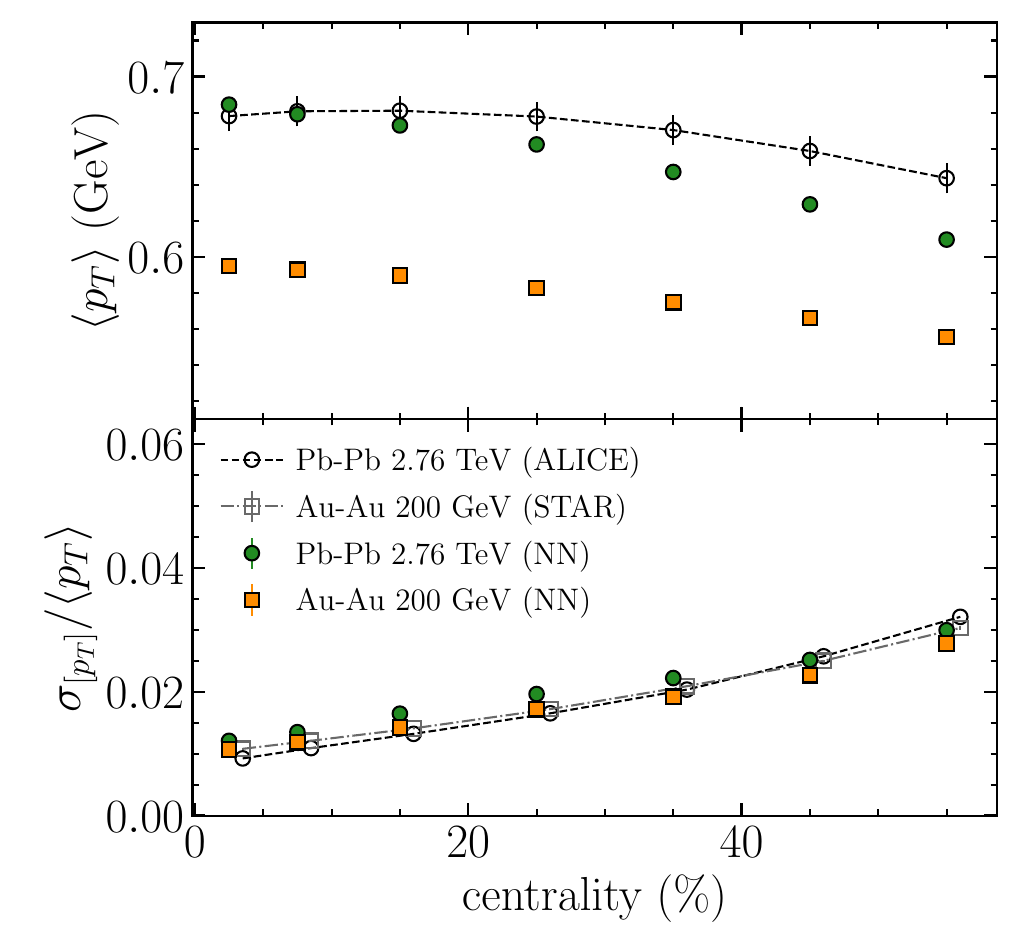}
    \caption{(Top) Mean transverse momentum $\media{p_T}$ of charged particles with $p_T \ge 0.15$~GeV, $|\eta|<0.5$, as a function of centrality, in 10\%-wide classes (0--80\%) defined from the multiplicity predicted by the NN, for Pb--Pb collisions at $\sqrt{s_{NN}} = 2.76$~TeV and Au--Au  collisions at $\sqrt{s_{NN}} = 200$~GeV, compared with ALICE~\cite{ALICE:2012aqc} data ($0.15\le p_T\le5$~GeV, $|\eta|<0.8$). (Bottom) Relative fluctuations $\sigma_{[p_T]}/\media{p_T}$ compared with ALICE~\cite{ALICE:2014gvd} ($0.15\le p_T\le2$~GeV, $|\eta|<0.8$) and STAR~\cite{STAR:2005vxr}. Error bars on the model results represent jackknife statistical uncertainties. }
    \label{fig:pt}
\end{figure}

Figure~\ref{fig:pt} (top) shows $\media{p_T}$ of charged particles as a function of centrality for both systems, compared with ALICE data~\cite{ALICE:2012aqc}.  There is no STAR result with this $p_T$ cut. Since Pb and Au nuclei have similar sizes, the difference between the two systems mainly reflects the difference in collision energy, the higher energy density reached at the LHC results in a larger $\media{p_T}$. However, the model does not fully describe the Pb--Pb data, deviating by up to 5--6\% for centralities between 50\% and 70\%. Note that Ref.~\cite{JETSCAPE:2020mzn} calibrated only the $\media{p_T}$ of identified pions, kaons, and protons; the $\media{p_T}$ of all charged particles was not included in the Bayesian analysis.

Figure~\ref{fig:pt} (bottom) shows the relative fluctuations $\sigma_{[p_T]}/\media{p_T}$ for both systems, compared with ALICE data for Pb--Pb~\cite{ALICE:2014gvd} and STAR data for Au--Au~\cite{STAR:2005vxr}. Statistical uncertainties on $\sigma_{[p_T]}/\media{p_T}$ and on the Pearson
correlation coefficients discussed below are estimated with the jackknife
method. Experimentally, the dynamical fluctuations are extracted from the two-particle correlator $C_m = \langle \Delta p_{T,i}\,\Delta p_{T,j} \rangle$, which removes the statistical fluctuations due to the finite number of particles. Since the NN predicts the mean response to each initial condition, the fluctuations of the predicted $[p_T]$ contain no statistical component and can be compared directly with $\sqrt{C_m}/\media{p_T}$, up to non-flow contributions such as resonance decays and jets, which are not included in our simulation. The ALICE data in the bottom panel use $0.15 \le p_T \le 2$~GeV~\cite{ALICE:2014gvd}, narrower than the $0.15\le p_T \le5$~GeV range of the model; this mismatch, of the same kind that leads us to decline a direct data comparison for $\rho(N_{\rm ch},[p_T])$ in eq. \eqref{eq:rho2}, is expected to affect the absolute scale of $\sigma_{[p_T]}/\media{p_T}$, since the upper cut changes the dynamic range of $[p_T]$. With this caveat, our results agree with the STAR data for Au--Au across the full centrality range. For Pb--Pb, the agreement with ALICE is good at mid-central and peripheral centralities, but the predicted fluctuations exceed the data by 20--30\% in the most central classes; since ALICE selects these classes with the V0M estimator, part of this difference may also reflect the centrality-estimator mismatch that is the theme of this paper, rather than only the $p_T$-range mismatch. At mid-central and peripheral centralities, the relative fluctuations of the two systems are also close to each other at the same centrality despite the different multiplicities; this approximate equality is weaker in the most central classes.

This behavior has a simple interpretation. The ratio $\sigma_{[p_T]}/\media{p_T}$ is independent of the overall momentum scale, which depends on the temperature reached in the collision. In hydrodynamics, $[p_T]$ responds approximately linearly to fluctuations of the initial state~\cite{Bozek:2012fw, Bozek:2017elk, Schenke:2020uqq}, so the relative fluctuations of $[p_T]$ are proportional to the relative initial-state fluctuations, with a response coefficient that depends on the equation of state and transport properties of the medium~\cite{Gardim:2020sma}.
At the same centrality, Au--Au and Pb--Pb collisions have a similar number of participant nucleons and, in our model, the same initial-state parameters, and hence comparable initial-state fluctuations. The approximate equality of $\sigma_{[p_T]}/\media{p_T}$ at the two energies in the mid-central and peripheral region therefore suggests that this response coefficient depends only weakly on collision energy there; in the most central classes, where the two systems differ by up to 15\%, this weak dependence does not hold as well. This comparison is possible only because of the $1.5\times10^6$ events generated per system. With the event samples of typical hybrid simulations, both the centrality determination and the statistical noise would obscure the comparison.


\section{Correlations of the Mean Transverse Momentum with Elliptic Flow and Multiplicity}
\label{rho}

Although anisotropic flow was not included in the training of the present version of GOUAI, the correlation between elliptic flow and $[p_T]$ can still be estimated. Elliptic flow responds approximately linearly to the initial eccentricity, $v_2 = k_2\varepsilon_2$~\cite{Gardim:2011xv,Niemi:2012aj}. Since the Pearson correlation coefficient is invariant under rescaling of either variable, the response coefficient $k_2$ cancels in the correlator proposed in Ref.~\cite{Bozek:2016yoj}, provided it does not fluctuate from event to event within a centrality class:

\begin{align}
\rho(v_2^2, [p_T]) &= \frac{\operatorname{cov}(v_2^2,[p_T])}
{\sigma_{v_2^2}\,\sigma_{[p_T]}} \approx
\frac{\operatorname{cov}(\varepsilon_2^2,[p_T])}
{\sigma_{\varepsilon_2^2}\,\sigma_{[p_T]}} , \\
\rho(N_{\rm ch}, [p_T]) &= \frac{\operatorname{cov}(N_{\rm ch},[p_T])}
{\sigma_{N_{\rm ch}}\,\sigma_{[p_T]}} ,
\label{eq:rho2}
\end{align}

where $\operatorname{cov}(a,b) = \langle ab \rangle - \langle a \rangle\langle b \rangle$ and $\sigma_a^2 = \operatorname{cov}(a,a)$, with averages taken over the events in each centrality class. This approximation has been shown to capture the main features of full hydrodynamic calculations~\cite{Giacalone:2020dln}, although it neglects the dependence of $k_2$ on system size, which is itself correlated with $[p_T]$, as well as nonlinear response in peripheral collisions.

\begin{figure}[h]
    \centering
    \includegraphics[width=1.0\linewidth]{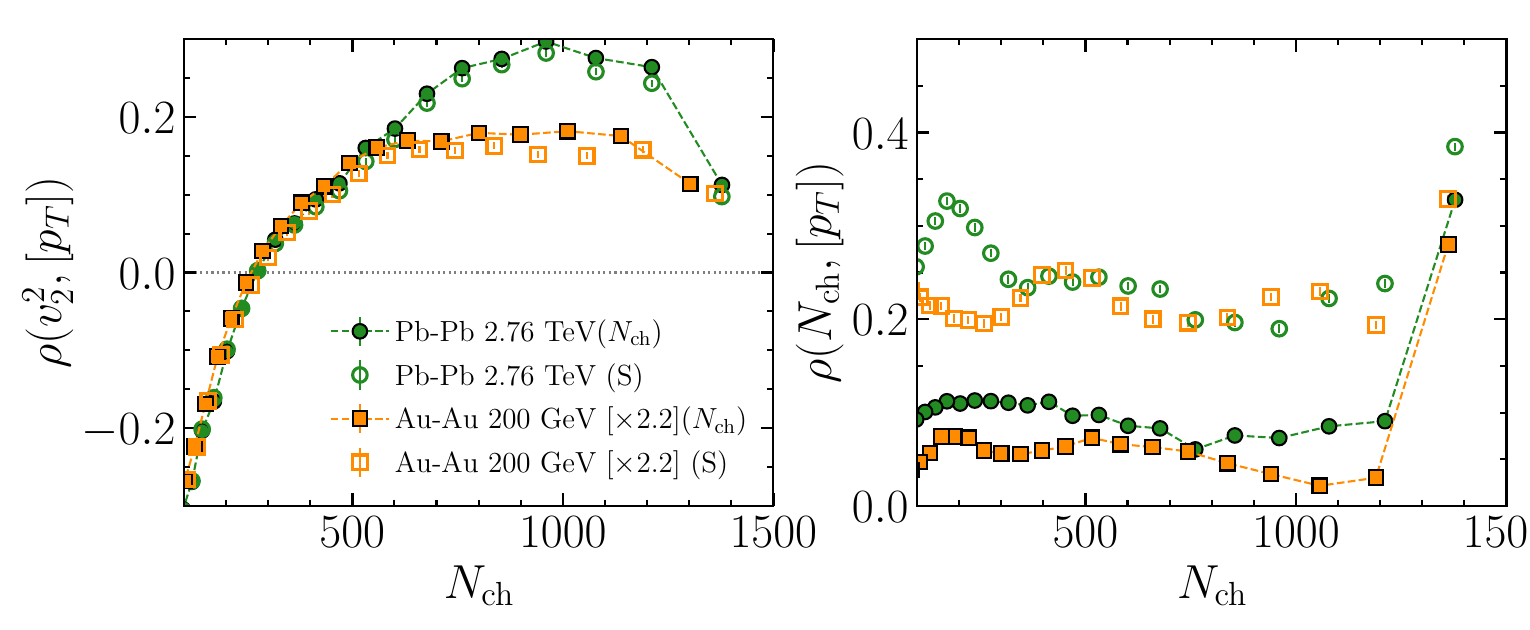}
    \caption{Pearson correlation coefficients $\rho(v_2^2, [p_T])$, computed with $\varepsilon_2^2$ in place of $v_2^2$ (left, Eq.~\eqref{eq:rho2}), and $\rho(N_{\rm ch}, [p_T])$ (right), for Pb--Pb collisions at $\sqrt{s_{NN}} = 2.76$~TeV and Au--Au collisions at
$\sqrt{s_{NN}} = 200$~GeV, as functions of the charged-particle multiplicity, with the Au--Au multiplicity scaled by a factor 2.3 (see text). Filled symbols: centrality defined from $N_{\rm ch}$. Open symbols: centrality defined from the initial entropy $S$; events are still binned and plotted against their multiplicity, not $S$, so that both definitions share the same abscissa. Observables are computed for charged particles with $p_T \ge 0.15$~GeV, $|\eta|<0.5$, in 3\%-wide centrality classes. Error bars represent jackknife statistical uncertainties.}
    \label{fig:rho}
\end{figure}

We computed $\rho(v_2^2, [p_T])$ and the correlation between multiplicity and mean transverse momentum, $\rho(N_{\rm ch}, [p_T])$, for both systems, in 3\%-wide centrality classes, using two centrality definitions: the final-state multiplicity $N_{\rm ch}$, our closest proxy to the experimental estimators, and the initial entropy, which is accessible only in
simulations. Both correlators are computed from the network's mean-response predictions, which contain no finite-particle statistical noise. The results are shown in Fig.~\ref{fig:rho}. To compare the two systems, the multiplicity of Au--Au collisions is scaled by a factor 2.3, the ratio of the mean charged-particle multiplicities of Pb--Pb and Au--Au collisions in the [0-5\%] centrality class. Since this rescaling maps events of similar centrality onto the same multiplicity range, it provides a comparison at approximately equal centrality. The correlator $\rho(v_2^2, [p_T])$ is essentially insensitive to the centrality definition, consistent with the fact that neither correlated quantity is the variable used to define centrality. It is similar for both systems over most of the centrality range, except between 10\% and 20\%, where Pb--Pb shows a markedly stronger correlation than Au--Au. We have not identified whether this localized difference reflects genuine physics or a limitation of the current networks.

In contrast, $\rho(N_{\rm ch}, [p_T])$ depends strongly on the centrality definition, see fig. \ref{fig:rho}. With $N_{\rm ch}$-based centrality, it is small and positive, and consistent between Pb--Pb and Au--Au over the full centrality range. With entropy-based centrality, it is substantially larger and positive for both systems, which track each other closely in shape. The origin of this difference is that, with $N_{\rm ch}$-based centrality, the range of $N_{\rm ch}$ within each class is restricted by construction, which suppresses its correlation with
any other quantity, whereas with entropy-based centrality $N_{\rm ch}$ fluctuates freely within each class. Under entropy-based centrality, $\rho(N_{\rm ch},[p_T])$ is essentially the correlation between the two predicted residuals, $d_{N_{\rm ch}}$ and $d_{[p_T]}$. Since these share a ResNet trunk and a common dense layer, we checked that this correlation is not an artifact of the shared representation by computing it directly from the hybrid simulations, on the $\sim600$ events used for training and validation: $\rho(d_{N_{\rm ch}},d_{[p_T]}) = 0.48$ (Pb--Pb) and $0.46$ (Au--Au) in the hybrid model, compared with $0.53$ and $0.49$ from the network on the same events, consistent in sign and magnitude. The size of this effect depends on the class width, since a wider class allows a larger variation of $N_{\rm ch}$, as already pointed out for correlations between flow harmonics~\cite{Gardim:2016nrr}.

This result has practical consequences. Many theoretical studies define centrality from the initial state because it is inexpensive and exactly known in simulations, whereas experiments must use a final-state observable. Our results show that $\rho(v_2^2, [p_T])$ is nearly insensitive to this choice, whereas $\rho(N_{\rm ch}, [p_T])$ is not, for which the centrality definition changes the magnitude and centrality dependence of the prediction substantially, even though the two collision systems remain compatible with each other under either definition. This observable is closely related to the correlation between $\media{p_T}$ and $N_{\rm ch}$ in ultracentral collisions used to extract the speed of sound of the QGP~\cite{Gardim:2019xjs, CMS:2024sgx}, where the sensitivity to the centrality determination has also been noted~\cite{Nijs:2023bzv}. The relevant estimator is typically in a different acceptance than the multiplicity entering the correlator, which our single-acceptance midrapidity estimator does not reproduce. Predictions for correlators involving the multiplicity should therefore be
computed with an event selection that emulates the experimental estimator in both definition and acceptance, as closely as the model allows.

A direct comparison with data was not possible, since the available measurements were performed at a different collision energy, Pb--Pb, or with a different transverse-momentum range, Au--Au.

\section{Conclusion}

A meaningful comparison between hydrodynamic models and experimental data requires centrality to be defined in the same way in both, from the final-state multiplicity. In simulations, this has so far been impractical, since it requires millions of hybrid events, and centrality has instead been approximated using the initial entropy or energy. In this work, we have shown that a neural network trained on only a few hundred fully simulated events of a state-of-the-art hybrid model can predict the event-by-event multiplicity $N_{\rm ch}$ and mean transverse momentum $[p_T]$ with relative errors of about 1--2\% for $N_{\rm ch}$ and about 1\% for $[p_T]$ up to 80\% centrality, making it possible to generate millions of events and to define centrality from a multiplicity-based estimator much closer to experiment than
initial-state proxies, though still not identical to any single experimental one (Sec.~\ref{NN:centrality}). The networks learn the event-by-event
deviations from the smooth dependence on the total initial entropy, which
is what distinguishes events of the same entropy, and their accuracy does not
depend on whether the initial condition is given as entropy or energy density.

Applying this approach, implemented in GOUAI, to Pb--Pb collisions at $\sqrt{s_{NN}} = 2.76$~TeV and Au--Au collisions at $\sqrt{s_{NN}} = 200$~GeV, we found that centrality classes defined from the initial entropy differ from those defined from the final multiplicity, and the difference grows as the classes become narrower. In the 5--10\% class, the two definitions agree for 96--97\% of the events in both systems, falling to 83--95\% in 1\%-wide classes, with migration only to adjacent classes. This is particularly relevant because correlations should be computed in narrow classes, typically 1\% wide, before being combined into wider ones~\cite{Gardim:2016nrr}.

With centrality defined from the multiplicity, the mean transverse momentum of charged particles in Pb--Pb collisions predicted with the maximum a posteriori parameters of Ref.~\cite{JETSCAPE:2020mzn} deviates from ALICE
data by up to 5--6\% for centralities between 50\% and 70\%. The relative fluctuations of $[p_T]$ agree with the STAR data for Au--Au across the full centrality range; for Pb--Pb the agreement with ALICE is good at mid-central and peripheral centralities but the model overpredicts the fluctuations by 20--30\% in the most central classes. In the mid-central and peripheral region, the fluctuations of both systems are close to each other at the same
centrality despite the different multiplicities, suggesting that they are driven mainly by initial-state geometric fluctuations, with a hydrodynamic response that depends only weakly on collision energy there; since both systems share identical initial-state parameters by construction in our model, this is only a partial test of that universality.

We also computed the Pearson correlation coefficients $\rho(v_2^2, [p_T])$, estimating elliptic flow from the linear response to the initial eccentricity, and $\rho(N_{\rm ch}, [p_T])$. The former is practically insensitive to the centrality definition and similar for both systems over most of the centrality range, with a localized difference between the two systems between 10\% and 20\% centrality whose origin is not yet understood. The latter, in contrast, depends strongly on the centrality definition, with its magnitude and centrality dependence changing substantially between the two definitions, while remaining compatible between the two collision systems under either definition. Since this correlation is closely related to the observables used to extract the speed of sound of the QGP, predictions for it must be computed with an event selection that emulates the experimental estimator as closely as possible. This study is left for future work.

The two issues are connected: correlations must be computed in narrow classes to avoid spurious contributions from the variation of centrality within each class~\cite{Gardim:2016nrr}, and it is in narrow classes that the definition of centrality matters most.

These findings also have implications for Bayesian analyses, in which emulators are trained on a limited number of events per design point, and where the centrality definition alone can therefore affect the extracted parameters. Since the events already simulated at each design point are of the same
order as the training set used here, the approach presented here could in principle be extended to the full parameter space, providing large event samples and a consistent centrality definition at every design point. The added cost, dominated by training a network at each of the $O(500)$ design points, remains to be quantified against the Bayesian analysis itself, and
it is not yet established whether a few hundred events suffice across the full parameter space, where the structure of the residual may change with the model parameters, for instance the nucleon width. Natural next steps include incorporating anisotropic flow and the
fluctuations from particlization in the training, and applying the method at $\sqrt{s_{NN}} = 5.02$~TeV, where high-statistics data from LHC Runs 3 and 4 will allow centrality-dependent comparisons in very narrow classes.

\begin{acknowledgments}

The authors acknowledge the National Laboratory for Scientific Computing (LNCC/MCTI, Brazil), through the ambassador program (UNIFAL-MG), subproject STMHIC, for providing HPC resources of the SDumont supercomputer, which have contributed to the research results reported within this study url: http://sdumont.lncc.br, and the financial support from the Fundação de Amparo à Pesquisa do Estado de Minas Gerais (FAPEMIG) – Brazil APQ-00544-23.  FGG is supported by CNPq (Conselho Nacional de Desenvolvimento Cientifico) through 405458/2025-8 and 307806/2025-1, and by INCT-FNA (Instituto Nacional de Ciência e Tecnologia, Física Nuclear e Aplicações), research project No. 408419/2024-5.

\end{acknowledgments}

\bibliography{ref}   

\section*{Supplemental Material}
\setcounter{figure}{0}\renewcommand{\thefigure}{S\arabic{figure}}
\setcounter{equation}{0}\renewcommand{\theequation}{S\arabic{equation}}

\section*{Network, Training and Validation}

\subsection*{Training data}

For each system, the hybrid model described in the main text was run for 619 (Pb--Pb) and 584 (Au--Au) \TRENTo\ initial conditions spanning the full centrality range, with roughly equal numbers of events per 10\% class, about twice as many in 0--10\%, and few beyond 90\%. For each event we record the charged-particle multiplicity $N_{\rm ch}$ and mean transverse momentum $[p_T]$ at midrapidity for charged particles with $p_T \ge 0.15$~GeV. The events are split, stratified in $[p_T]$, into 450 (Pb--Pb) and 449 (Au--Au) training events and 169 and 135 held-out events. The held-out events are used only for the final evaluation (Fig.~\ref{fig:validation} of the main text and Figs.~\ref{figS:residual} and \ref{figS:centrality} here); they never enter training, early stopping, or model selection. The training events are further divided into five cross-validation folds of about 90 events, stratified in the total initial entropy.

\subsection*{Inputs}

Each \TRENTo\ energy-density profile, on a $280\times280$ grid with spacing 0.1~fm, is converted to entropy density $s$ with the lattice equation of state~\cite{HotQCD:2014kol} by cubic interpolation of the tabulated $s(\epsilon)$, with $s=0$ below the lowest tabulated energy density. The network receives two inputs: a two-channel image, whose first channel is the per-pixel entropy density divided by 255 (an arbitrary fixed scale, well above the maximum pixel value in the training set, that keeps the absolute-scale channel in the 8-bit image range conventionally used by the convolutional backbone) and whose second channel is the entropy density divided by the total entropy $S=\sum_i s_i$ (the geometry); and the scalar $S/S_{\rm max}$, where $S_{\rm max}$ is the largest total entropy in the training set.

\subsection*{Targets: trend and residual}

Both observables are dominated by a smooth dependence on $S$: a fit of $\log N_{\rm ch}$ as a function of $\log S$ alone explains more than 99\% of its variance, while $N_{\rm ch}$ itself spans three orders of magnitude. A network trained directly on $N_{\rm ch}$ spends its loss on reproducing this trend and has little incentive to learn the few-percent event-by-event deviations around it, which are the information that distinguishes events of equal $S$. We therefore fix the trend and train the network on the deviations only. For each observable $y \in \{N_{\rm ch}, [p_T]\}$ we write
\begin{equation}
y = g(S)\, e^{d}, \qquad \log g(S) = \sum_{k=0}^{3} a_k \left[\log(S/S_{\rm ref})\right]^k ,
\end{equation}
where $S_{\rm ref}$ is the median training $S$ and the coefficients $a_k$ are fitted by least squares to the training events of each fold and then kept fixed. Outside the range of the training data, $\log g$ is continued linearly. The network predicts the standardized residual
\begin{equation}
z = \frac{d-\mu}{\sigma}, \qquad d = \log y - \log g(S),
\end{equation}
with $\mu$ and $\sigma$ the mean and standard deviation of $d$ over the training events, separately for $N_{\rm ch}$ and $[p_T]$. The prediction is $y = g(S)\exp(\mu + \sigma z)$. Standardizing gives both targets the same weight in the loss and makes them homoscedastic. The last layer of each output head is initialized to zero, so that the untrained network reproduces $g(S)$ exactly and training can only add information beyond it.

\subsection*{Architecture}

The same architecture is used for both systems and both observables. The image branch is a ResNet-50 initialized with random weights, followed by a spatial-attention layer, a dilated $3\times3$ convolution with 512 filters, global average pooling, and dense layers of 1024 (with dropout 0.3) and 128 units. The scalar branch has dense layers of 64, 32, and 16 units. The two branches are concatenated into a shared dense layer of 128 units, which feeds two output heads, one for $z_{N_{\rm ch}}$ and one for $z_{[p_T]}$, each consisting of a dense layer of 32 units with dropout 0.4 and a linear output, both with $L_2$ regularization ($10^{-4}$). The loss is the sum of the Huber losses of the two standardized residuals.

\subsection*{Training and model selection}

Each training event is used together with twelve augmented copies: eight with random horizontal and vertical reflections and rotations of up to $\pm18^\circ$, and four in which, in addition, the geometry channel is transformed independently, with rotations of up to $\pm54^\circ$ and shifts of up to 1.4~fm. The networks are trained with the Adam optimizer (learning rate $3\times10^{-4}$, batch size 256). In every fold, the validation loss fluctuates strongly during roughly the first 150 epochs before stabilizing, as is common when a deep network is trained from scratch on a few hundred events. Neither early stopping, learning-rate reduction, nor checkpoint selection acts during the first 100 epochs. Afterwards, the learning rate is halved after 15 epochs without improvement of the validation loss, training stops after 80 such epochs, and the weights of the epoch with the lowest validation loss are kept; this happened between epochs 160 and 230 in all folds. One network is trained per cross-validation fold, on the other four folds. Each fold network predicts the events of its own fold, which it has never seen, so every training event receives an out-of-fold prediction, and the five networks agree within about 1\% for $N_{\rm ch}$ and 0.6\% for $[p_T]$ up to 80\% centrality. The final prediction is the geometric mean of the five fold networks, which on the held-out events is more accurate than a single network retrained on all training events.

\subsection*{Production}

For the $1.5\times10^6$ events per system, \TRENTo\ is run with the same parameters as for the hybrid simulations and the profiles are processed exactly as the training images, including the same equation-of-state conversion and normalization constants. About $10^{-5}$ of the events, grazing collisions with two participants, have an energy density everywhere below the lowest tabulated value and therefore zero total entropy; they are assigned $N_{\rm ch}=0$. The total entropy of 5\% (Pb--Pb) and 10\% (Au--Au) of the events lies below the smallest value in the training set; these are the most peripheral events, for which the predictions rely on the extrapolated trend.

\subsection*{Learning beyond the entropy trend}

\begin{figure*}[t]
    \centering
    \includegraphics[width=\textwidth]{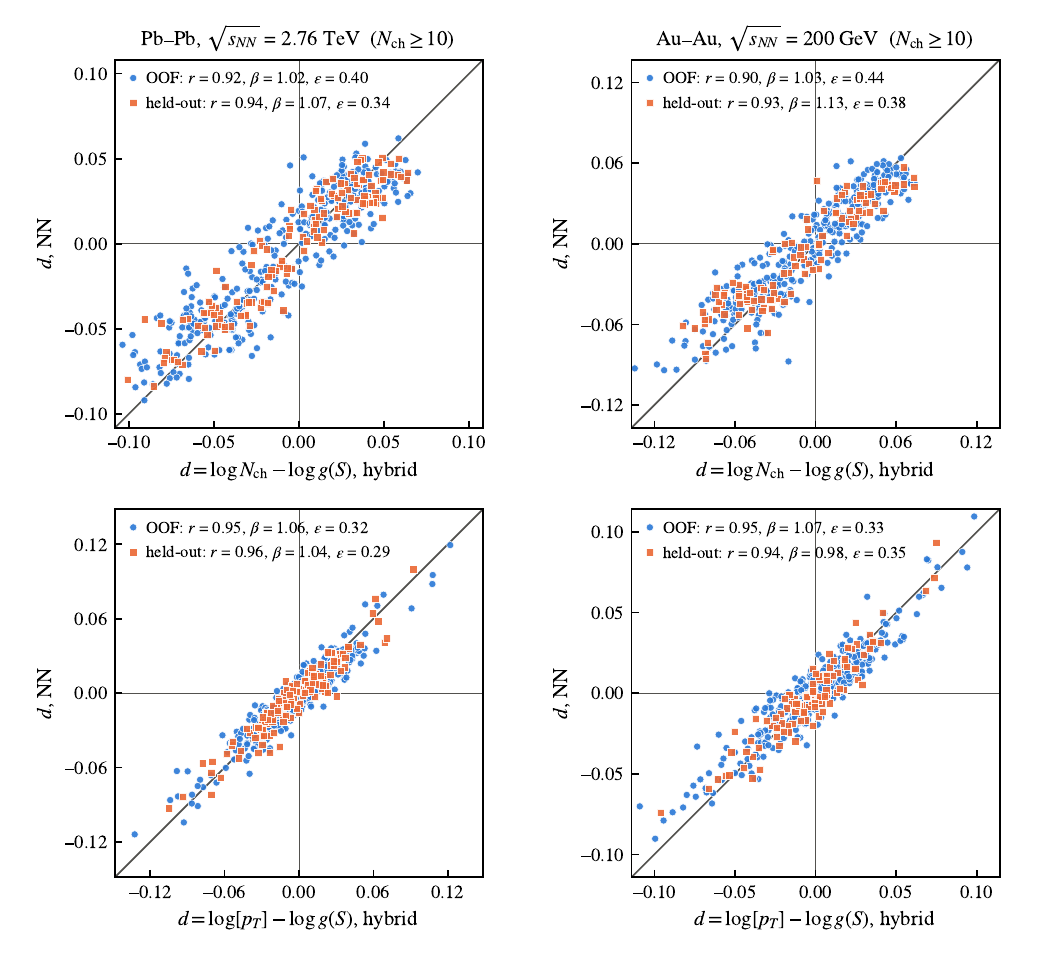}
    \caption{Event-by-event residual beyond the entropy trend, $d = \log y - \log g(S)$, predicted by the network versus the hybrid simulation, for $N_{\rm ch}$ (top) and $[p_T]$ (bottom), for Pb--Pb (left) and Au--Au (right) collisions, for events with $N_{\rm ch}\ge10$. Circles (OOF): out-of-fold predictions of the training events (382 Pb--Pb, 375 Au--Au). Squares: held-out events (145 and 113). The diagonal corresponds to a perfect prediction; a horizontal line at zero would correspond to the trend $g(S)$ alone. The legends give the correlation coefficient $r$, the calibration slope $\beta=\mathrm{cov}(d_{\rm hybrid},d_{\rm NN})/\mathrm{var}(d_{\rm NN})$, and $\epsilon$, the root-mean-square difference $d_{\rm NN}-d_{\rm hybrid}$ relative to the standard deviation of $d_{\rm hybrid}$.}
    \label{figS:residual}
\end{figure*}

Figure~\ref{figS:residual} shows the residual $d$ predicted by the network versus the true one, for events with $N_{\rm ch}\ge10$, both out-of-fold (training events, each predicted by a network that did not see it) and on the held-out events. For $N_{\rm ch}$, the spread of $d$ around the trend is about 4\%, and the network recovers it with a correlation coefficient $r = 0.90$--$0.94$; the root-mean-square error of the predicted residual is 0.34--0.44 of its spread, i.e., the network reduces the deviation from the trend $g(S)$ by a factor of 2.3--2.9. For $[p_T]$, $r = 0.94$--$0.96$ and the error is 0.29--0.35 of the spread. The calibration slope $\beta=\mathrm{cov}(d_{\rm true},d_{\rm pred})/\mathrm{var}(d_{\rm pred})$ is close to one (0.98--1.13). The out-of-fold and held-out results agree, showing no sign of overfitting.

\subsection*{Accuracy per centrality class}

\begin{figure*}[t]
    \centering
    \includegraphics[width=\textwidth]{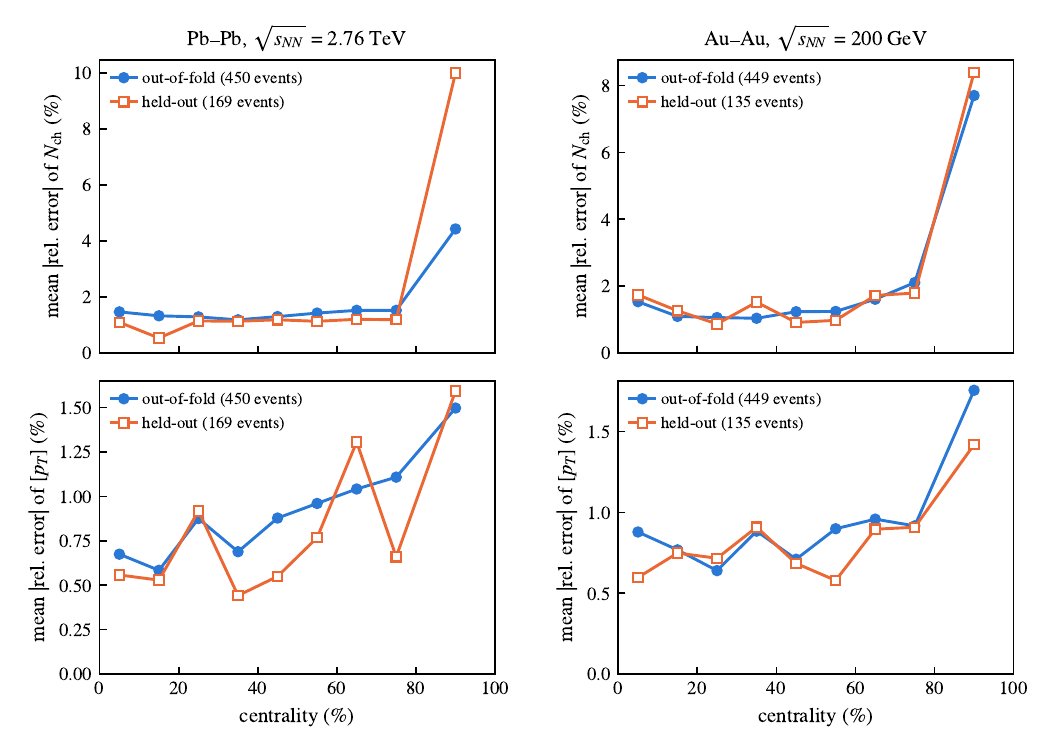}
    \caption{Mean absolute relative error of $N_{\rm ch}$ (top) and $[p_T]$ (bottom) per centrality class, for Pb--Pb (left) and Au--Au (right) collisions. Circles: out-of-fold predictions of the training events (450 Pb--Pb, 449 Au--Au). Squares: held-out events (169 and 135). Centrality is defined from the initial entropy.}
    \label{figS:centrality}
\end{figure*}

Figure~\ref{figS:centrality} shows the mean relative error in classes of centrality defined from the initial entropy. Up to 80\% centrality, the error is between 0.5\% and 1.8\% for $N_{\rm ch}$ on the held-out events (at most 2.1\% out-of-fold) and below 1.3\% for $[p_T]$ in both systems; the most central class, 0--10\%, shows only a mild increase (1.1--1.7\% for $N_{\rm ch}$). In the most peripheral class, 80--100\%, the $N_{\rm ch}$ error rises to 8--10\% on the held-out events (4--8\% out-of-fold), and the $[p_T]$ error to 1.4--1.6\% on the held-out events (up to 1.8\% out-of-fold). These events have $N_{\rm ch}\lesssim10$, where small absolute errors are large in relative terms, and they are sparsely represented in the training sample (5 Au--Au and 35 Pb--Pb training events beyond 90\% centrality); the five fold networks also disagree more there, by 5--6\% for $N_{\rm ch}$.

\subsection*{Entropy versus energy density as input}

\begin{figure*}[t]
    \centering
    \includegraphics[width=\textwidth]{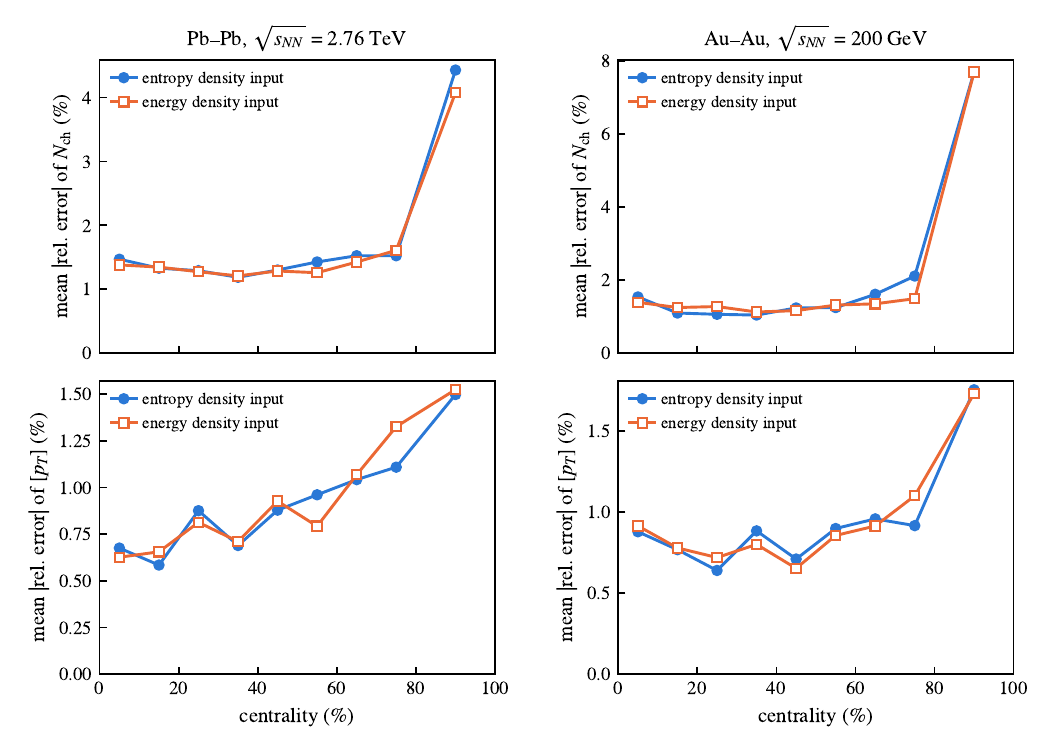}
    \caption{Mean absolute relative error of $N_{\rm ch}$ (top) and $[p_T]$ (bottom) per centrality class, out-of-fold, for networks with entropy density (circles) and energy density (squares) as input, trained on the same events with the same cross-validation folds, for Pb--Pb (left) and Au--Au (right) collisions.}
    \label{figS:inputs}
\end{figure*}

To test whether the choice of input affects the predictions, we trained a second set of networks identical in every respect, with the same training events, cross-validation folds, architecture, and training procedure, but with the \TRENTo\ energy density as input and the total energy in place of $S$. Because the two sets of networks predict exactly the same events under the same conditions, their per-event errors can be compared in pairs. Over the out-of-fold predictions of all training events with $N_{\rm ch}\ge10$, the mean relative error of $N_{\rm ch}$ is 1.39\% (entropy) versus 1.36\% (energy) for Pb--Pb and 1.38\% versus 1.33\% for Au--Au; for $[p_T]$ it is 0.97\% versus 0.99\% and 1.01\% versus 1.01\%. None of these differences is significant: the bootstrap 95\% confidence intervals of the mean paired difference include zero, and Wilcoxon signed-rank tests give $p\ge0.4$. The same holds on the held-out events. Per centrality class (Fig.~\ref{figS:inputs}), the differences are small and of either sign: 3 of the 36 class-by-class comparisons (9 classes, 2 observables, 2 systems) are nominally significant at the 5\% level, two favoring energy and one favoring entropy, consistent with chance. The two networks' predictions for the same event differ on average by 0.8--0.9\% for $N_{\rm ch}$ and 0.5\% for $[p_T]$, less than their individual errors. The initial-state total alone is also an equally good proxy in both cases: the spread of $\log N_{\rm ch}$ around its trend in the total energy is even slightly smaller than in the total entropy. Within the accuracy of the networks, entropy and energy density are therefore equivalent inputs.

\end{document}